\documentclass[twocolumn]{aastex62}

\graphicspath{{./}{figures/}}
\usepackage{graphicx,subfigure}

\received{}
\revised{}
\accepted{}

\shorttitle{Chemical complexity in turbulent regions}
\shortauthors{Cassone et al.}

\begin{document}
\title{Dust Motions in Magnetized Turbulence: Source of Chemical Complexity}
\correspondingauthor{Giuseppe Cassone,Cesare Cecchi-Pestellini}
\email{cassone@ibp.cz,cesare.cecchipestellini@inaf.it}

\author{Giuseppe Cassone$^\star$}
\affil{Institute of Biophysics of the Czech Academy of Sciences, Kr\'{a}lovopolsk\'{a} 135, 61265 Brno, Czech Republic}
\author{Franz Saija}
\affil{CNR-IPCF, Viale Ferdinando Stagno d'Alcontres 37, 98158 Messina, Italy}
\author{Jiri Sponer}
\affil{Institute of Biophysics of the Czech Academy of Sciences, Kr\'{a}lovopolsk\'{a} 135, 61265 Brno, Czech Republic}
\author{Judit E. Sponer}
\affil{Institute of Biophysics of the Czech Academy of Sciences, Kr\'{a}lovopolsk\'{a} 135, 61265 Brno, Czech Republic} 
\author{Martin Ferus}
\affil{J. Heyrovsky Institute of Physical Chemistry, Czech Academy of Sciences, Dolejskova 3, 18223, Prague 8, Czech Republic}
\author{Miroslav Krus}
\affil{Institute of Plasma Physics, Czech Academy of Sciences, Za Slovankou 1782/3, 18200 Prague, Czech Republic}
\author{Angela Ciaravella}
\affil{INAF {\textendash} Osservatorio Astronomico di Palermo, Piazza del Parlamento 1, 90134 Palermo, Italy}
\author{Antonio Jim\'enez-Escobar}
\affil{INAF {\textendash} Osservatorio Astronomico di Palermo, Piazza del Parlamento 1, 90134 Palermo, Italy}
\author{Cesare Cecchi-Pestellini$^\star$}
\affil{INAF {\textendash} Osservatorio Astronomico di Palermo, Piazza del Parlamento 1, 90134 Palermo, Italy}

\begin{abstract}
Notwithstanding manufacture of complex organic molecules from impacting cometary and icy planet surface analogues is well-established, dust grain-grain collisions driven by turbulence in interstellar or circumstellar regions may represent a parallel chemical route toward the shock synthesis of prebiotically relevant species. 

Here we report on a study, based on the multi-scale shock-compression technique combined with \emph{ab initio} molecular dynamics approaches, where the shock-waves-driven chemistry of mutually colliding isocyanic acid (HNCO) containing icy grains has been simulated by first-principles. At the shock wave velocity threshold triggering the chemical transformation of the sample ($7$~km~s$^{-1}$), formamide is the first synthesized species representing thus the spring-board for the further complexification of the system. In addition, upon increasing the shock impact velocity, formamide is formed in progressively larger amounts. More interestingly, at the highest velocity considered ($10$~km~s$^{-1}$), impacts drive the production of diverse carbon-carbon bonded species. In addition to glycine, the building block of alanine (i.e., ethanimine) and one of the major components of a plethora of amino-acids including, e.g., asparagine, cysteine, and leucine (i.e., vinylamine) have been detected after shock compression of samples containing the most widespread molecule in the universe (H$_2$) and the simplest compound bearing all the primary biogenic elements (HNCO). 

The present results indicate novel chemical pathways toward the chemical complexity typical of interstellar and circumstellar regions. 
\end{abstract}


\section{Introduction} \label{sec:intro}
Interstellar turbulence, because it is generally supersonic, creates in the interstellar medium a texture of low-velocity shocks and localized intense vortices \citep{HF12}, which may affect dust evolution more frequently and more significantly than the faster supernovae shock waves. Being cycled continuously through a variety of physical conditions, dust grains experience growth mechanisms and processes that redistribute grain mass into units of smaller size, or even entirely remove the solid component. 

Relative grain-grain motions arising from magnetohydrodynamic (MHD) turbulence are discussed by e.g., \citet{Y04}. 
The turbulent acceleration is modelled as the acceleration due to a spectrum of MHD waves, decomposed into incompressible Alfv\'enic modes and compressible magnetosonic modes. While the fluid motions accelerate grains through hydrodynamic drag due to the frictional interaction with the gas, electromagnetic fluctuations provide energy exchange involving resonant interactions between the particles and the waves, such as gyroresonance \citep{Y03} and transit accelerations. In particular, the gyroresonance mechanism can accelerate grains to supersonic speed relative to the gas. 

The effects of supersonic motions on dust grains may be relevant for chemistry, the most obvious example of this being the accumulation of ice mantles on the surfaces of dust grains. Since dust grains move fast through the turbulent interstellar gas, the accretion process may be powered, providing significant effects upon the distribution of molecular species both in the solid and gas phases \citep{Ge16}. Low-velocity grain collisions may also have dramatic chemical consequences by triggering grain mantle explosions (e.g., \citealt{Ca97,CP04,G11}). 

When two particles collide at sufficiently high velocity, strong shock waves are driven within them, compressing matter to very high pressures. Such event induces chemical variations of the initial constituents (\citealt{G10,M13}). If the colliding particles are covered with icy mantles, shock waves may significantly increase the complexity of the ice composition. Turbulence may also drive fragmentation, erosion and shattering (e.g., \citealt{Ca97}). The threshold velocities for grain-grain shattering are of the order of 1~km~s$^{-1}$. Since a shattering event occurs over timescales comparable to the collision time (e.g., $\Delta t \approx 1$~ps for vaporization, \citealt{T94}), chemistry may be already relaxed, and the products of the chemical reactions taking place during the collisions are eventually ejected into the gas-phase.

Many organic molecules of moderate complexity, such as ethanol and glycolaldehyde, are detected at relatively high abundances in various interstellar locations, especially in regions of star formation \citep{HVD09,WV}, and are considered to be related to astrobiology. Such species cannot be readily formed by conventional interstellar gas-phase chemistry, and are thought to form through a chemistry involving dust grains. It is known that fairly simple molecular ices accumulating on the surfaces of dust grains in dense gas in star-forming regions transform into relatively complex gas phase species, apparently in some form of solid-state chemistry, involving various radicals trapped in the grain mantles, when activated by heating of the accreting central protostar (e.g., \citealt{Ga08,Ga13,Cu17}). All these processes can be replicated in laboratory experiments.

The dominant role of grain surface chemistry has been challenged observationally in cold environments (e.g., \citealt{M07,JS16}). Moreover, \citet{ER16} have theoretically suggested that combination of radicals trapped in amorphous water ice (the major component of interstellar ices) may not result in larger molecules. Other theoretical studies propose that complex organic molecules may be formed via suitable gas-phase reaction routes  (e.g., \citealt{K13,S17}). Still, laboratory experiments show that processing of interstellar ice analogues may lead to the formation of complex organic molecules via energetic (e.g., \citealt{MC02,O09,C13}) and non-energetic (e.g., hydrogenation of CO and of other small radicals, \citealt{F15,C16,F17}) routes, while recombination of radicals may occur on the fly, in the transient high-density gas-phase during non canonical mantle explosions \citep{R13}. 

Thus far, amino-acids have not been identified in the interstellar medium. However, a few species with the peptide moiety have been detected, e.g., formamide (NH$_2$CHO), perhaps the most important for proteins. The role played by formamide in the emergence of terrestrial life is one of the hottest subjects of contemporary research on the origins of life \citep{FS16,Sp16}. Formamide may serve as a universal feedstock molecule both for the one-pot \citep{SAL15} as well as for the multistep high-yield \citep{B16} synthesis of nucleosides. Such species may have also contributed  to the formation of nucleobases during extraterrestrial impacts on the early Earth \citep{FER15}. Moreover, {\it in silico} simulations of the Miller experiment revealed that formamide plays the role of a key intermediate in the reaction pathway \citep{SAI14}. While formamide is readily formed in interstellar ice analogues (e.g., \citealt{J11} and references therein), its gas-phase synthesis has been also suggested \citep{K13,B15}. Possible support to this new chemical scenario may come from the observations of formamide emission in a shocked region around a solar-type protostar \citep{C17}. However, \citet{Q18} have shown that either gas-phase formation or grain surface synthesis may dominate depending on the physical conditions of the source. So both formation routes may possibly co-exist.

Related to formamide is isocyanic acid (HNCO), formally the dehydrogenation product of the simplest amide, which -- along with its structural isomer cyanic acid -- represents the smallest stable molecule containing all four primary biogenic elements. This species has been observed in a variety of Galactic and extragalactic environments (e.g., \citealt{LS15} and references therein), as well as in processed interstellar icy analogues \citep{JE14,F15,K16}. A recent combined experimental-theoretical study has investigated the HNCO-based synthesis of formamide in exotic planetary atmospheres (\citealt{F18}).

In this work we explore the chemical processes relevant to colliding HNCO-containing icy grains, whose motions are driven by turbulence in inter- or circum-stellar regions. We simulate the event through uni-axial shock waves described by the multi-scale shock-compression simulation technique (MSST, \citealt{MSST}). The evolving chemistry is then followed exploiting \emph{ab initio} molecular dynamics (AIMD) approaches. 

\section{Model and Methods}
Laboratory simulations suggest that a very rich chemistry may occur in molecular ices on dust surfaces. The wide range of products hints at the operation of a radical-radical association chemistry, even though mechanistic details of the processing are still largely unknown. 

The formation of HNCO in ices proceeds through reaction between CO molecules and radical intermediates involved in the formation of NH$_3$, i.e. NH and NH$_2$ \citep{F15}. The formation of isocyanic acid competes with the formation of ammonia in non-polar CO ices, and becomes the favored channel when the atomic hydrogen accretion rate is slow. Ultraviolet photo-processing of water ices containing nitrogen-bearing species may lead to the HNCO synthesis \citep{JE14}. At low temperatures other energetic processings of nitrogen-containing solid mixtures of various compositions drive the synthesis of formamide together with isocyanic acid (e.g., \citealt{K16}).

When the densities are high enough for CO to freeze-out onto grains, molecular hydrogen is by far the most abundant hydrogen-bearing gas-phase species in dense regions. While H$_2$ is not expected to accrete firmly onto substrates at temperatures larger than $\approx 7$~K, it can be trapped in the micropores of CO ice. Upon investigating thermal- and photo-desorption of CO ices, \citet{MC10} observed that molecular hydrogen desorbed from the ice at the beginning of the warm-up process ($\approx 8$~K). The desorption reached a maximum at 14~K and was completed at 20~K. A possible interpretation for this observation is that H$_2$ is moving around by quantum tunneling in the amorphous CO ice. Thus, in cold regions H$_2$ would not only be transiently deposited on the ice surface, it would also accumulate in the bulk. Although in standard conditions H$_2$ molecules have a low reactivity compared to atomic hydrogen, during compressions this turns out not to be the case. 

The great advantage of computations when complementing experiments is that they provide information on selected single molecules and chemical reactions (e.g., \citealt{SPO16}, \citealt{C18}). In this work, we have deployed numerical calculations to follow the chemical evolution of a mixture of isocyanic acid and molecular hydrogen, when the constituent icy mantle is abruptly subjected to extreme pressures caused by grain-grain collisions in space conditions. A sample composed by $32$~HNCO and $50$~H$_{2}$ molecules (i.e., 228 atoms) is simulated by means of Density-Functional-Theory (DFT) based Born-Oppenheimer molecular dynamics exploiting the CP2K molecular simulation software \citep{Hu14}. The starting simulation boxes (i.e., the super-cells) have been prepared such that the internal pressure -- determined by means of first-principles evaluation of the stress tensor -- were equal to $5$~GPa. As usual, all the structures were replicated in space by means of periodic boundary conditions. We used a plane-wave cutoff of $400$~Ry and optimized triple zeta valence polarized basis set for all elements in the system. Goedecker-Teter-Hutter pseudopotentials~\citep{GTH1} along with the D3(BJ)~\citep{Grimme1,Grimme2} dispersion-corrected Perdew-Burke-Ernzerhof~\citep{P96} exchange and correlation functional have been employed. 

Shock waves are described by means of the MSST \citep{MSST}. This approach relieves typical system-size issues, at the same time allowing for a realistic description of chemistry under extreme conditions~\citep{G10}. Each molecular dynamics step, corresponding to 0.5 fs of dynamics, required about 100 s in order to be completed on a computer cluster exploiting -- for each simulation carried out -- 32 Intel ES 4650 processors and optimized inter-node ultra-fast communications (i.e. InfiniBand FDR10 Mellanox). After compression of the sample, a decompression of the simulation boxes enables the identification of the ``stable'' products stemming from the impact-induced chemical reactions. We have tested uni-axial shock wave velocities ranging from 6 to $10$~km~s$^{-1}$. This way, shock-compressed thermodynamic states identifiable with pressures $P_S = 37$~GPa ($6$~km~s$^{-1}$), $49$~GPa ($7$~km~s$^{-1}$), $58$~GPa ($8$~km~s$^{-1}$), $65$~GPa ($9$~km~s$^{-1}$), and $72$~GPa ($10$~km~s$^{-1}$) are generated through the MSST within dynamical time-scales of about $5-10$~ps. Then, a decompression process of the simulation boxes follows the maximally compressed state of each simulation cell, leading each system to a final, decompressed, pressure of about $5$~GPa, corresponding to the unperturbed (starting) simulation cell. All the chemical yields -- unless explicitly specified -- refer to the initial amount of HNCO molecules. 
\begin{figure}
\includegraphics[width=0.5\textwidth]{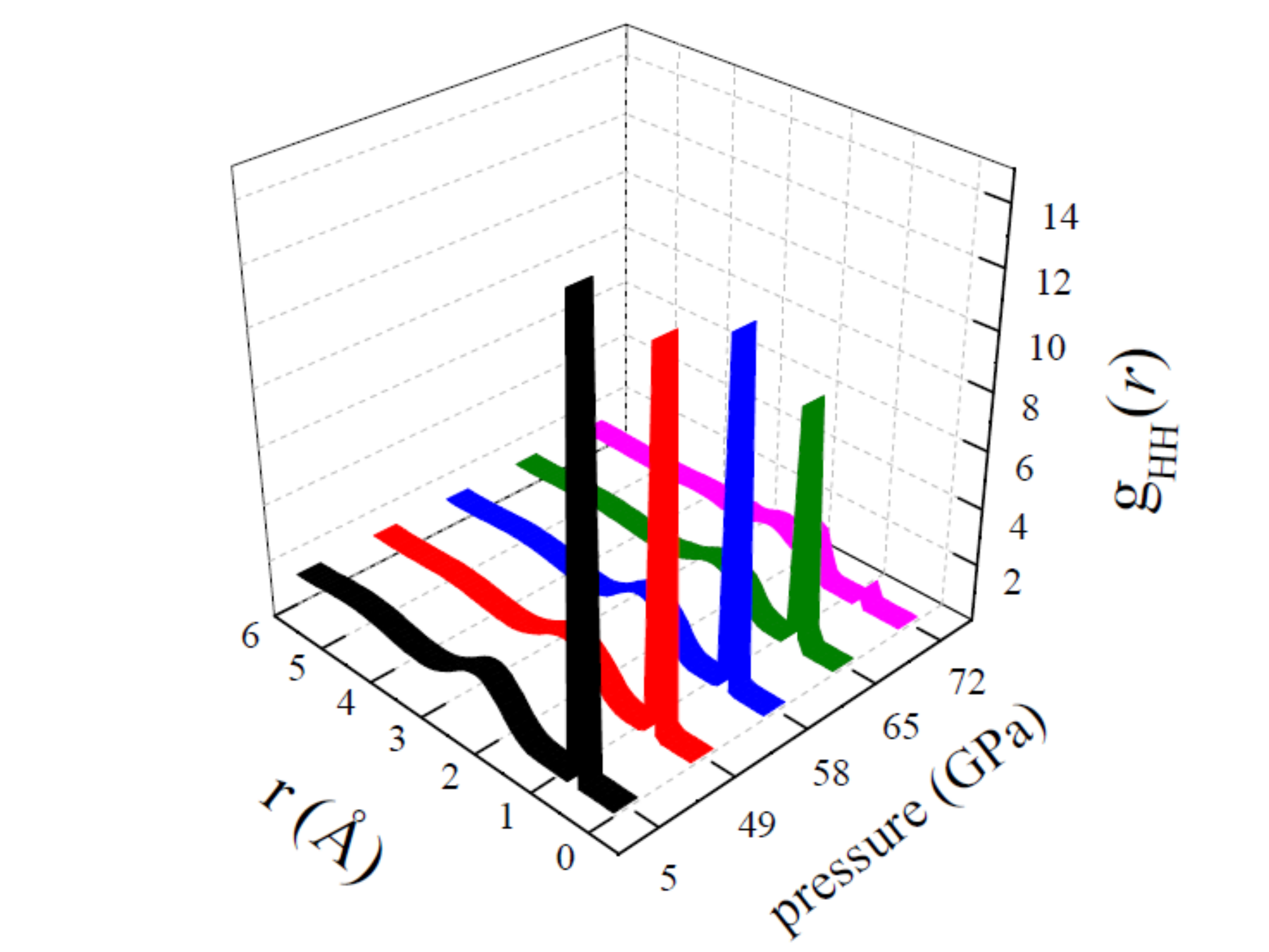}
\caption{Hydrogen-hydrogen RDF for different shock pressures, determined after decompression of the simulation boxes.}
\label{fone}
\end{figure}

Simulation of the global process of decompression as well as that of the ``decompressed'' states is performed for times longer than $20$~ps. These time-scales allow for the evaluation of the (meta)stable species in each system by means of direct inspection of the trajectories. Especially the atomic radial distribution functions (RDFs) have been very useful to follow the systematic increase of molecular complexity as a function of increasing shock wave velocities. RDFs describe how the density of a specific particle varies as a function of distance from a reference particle. More precisely, they represent the average densities of particles (atoms, molecules, etc.) at the position $r$, given that a tagged particle (atom, molecule, etc.) is at a pre-selected origin~\citep{AT97}; operationally, RDFs have been determined as three-dimensional histograms over the particles' labels averaged over time and space. Thus, numerical results are referred to the ``decompressed'' state of the simulation box, which is achieved after the relaxation of the compressed state. Finally, each decompression process starts from the nominal Hugoniot temperature of each given shock-compressed state and it is conducted at several temperatures in the range $300 - 2500$~K in order to check for eventual temperature-dependent chemical reactions occurring after each compressed state. It turned out that, notwithstanding a trivial enhancement of the molecular vibrations, no differences in the final composition of the products has been recorded for different temperature values proving -- a posteriori -- that all the chemical transformations are pressure-driven reactions. To this aim, distinct Car-Parrinello~\citep{CP} molecular dynamics simulations with the Parrinello-Rahman Lagrangian~\citep{PR} for the motion of the simulation box are performed by means of the plane-wave/pseudopotentials software package Quantum~ESPRESSO~\citep{G09} in order to independently reproduce the decompression process of the simulation super-cells. In these cases, the fictitious electronic masses are set to a value of $300$~a.u., with a cutoff energy of $40$~Ry, and a cutoff energy for the charge density of $320$~Ry, which allowed us to adopt a timestep of $0.12$~fs. In all cases, the dynamics of nuclei is simulated classically using the Verlet algorithm.

Although CO in icy mantles may in principle be involved in the overall chemistry here described, due to the substantial computational demand of the calculations (i.e., $\approx 350000$ CPU hours), we will consider the inclusion of this molecule in a future work. In this way, it will be possible, inter alia, to discern not only how the overall reaction network is affected by the presence of CO but also how each reaction pathway is inhibited/catalyzed by the presence of such a species. Additionally to these considerations, another rationale holds for the insertion of water molecules to the initial samples. In fact, CO ice in the interstellar medium is mainly found on top of pre-existing H$_2$O icy layers (e.g., \citealt{P08}), and hence it is expected to be shock-compressed on time-scales of the order $\Delta t \ga w/v_{\rm d}$, $w$ being the icy mantle thickness. For $v_{\rm d} = 10$~km~s$^{-1}$, and $w \approx 10$~nm, we obtain $\Delta t \ga 1$~ps, sizably longer than those characterizing the initial chemical activity that takes place in presence of HNCO and H$_2$ on the surface. Moreover, as it will be laid out in the next section, since water is copiously produced at the most intense shock regime, the latter species is somehow involved in the post-collision chemistry in the respective sample.

\section{Results}
Averaged molecular correlations can be quantified by means of the RDFs~\citep{HMcD}. Moreover, by sampling shorter atom-to-atom distances, important intra-molecular insights can be earned in determining these structural functions. 
\begin{figure}
\includegraphics[width=0.4\textwidth]{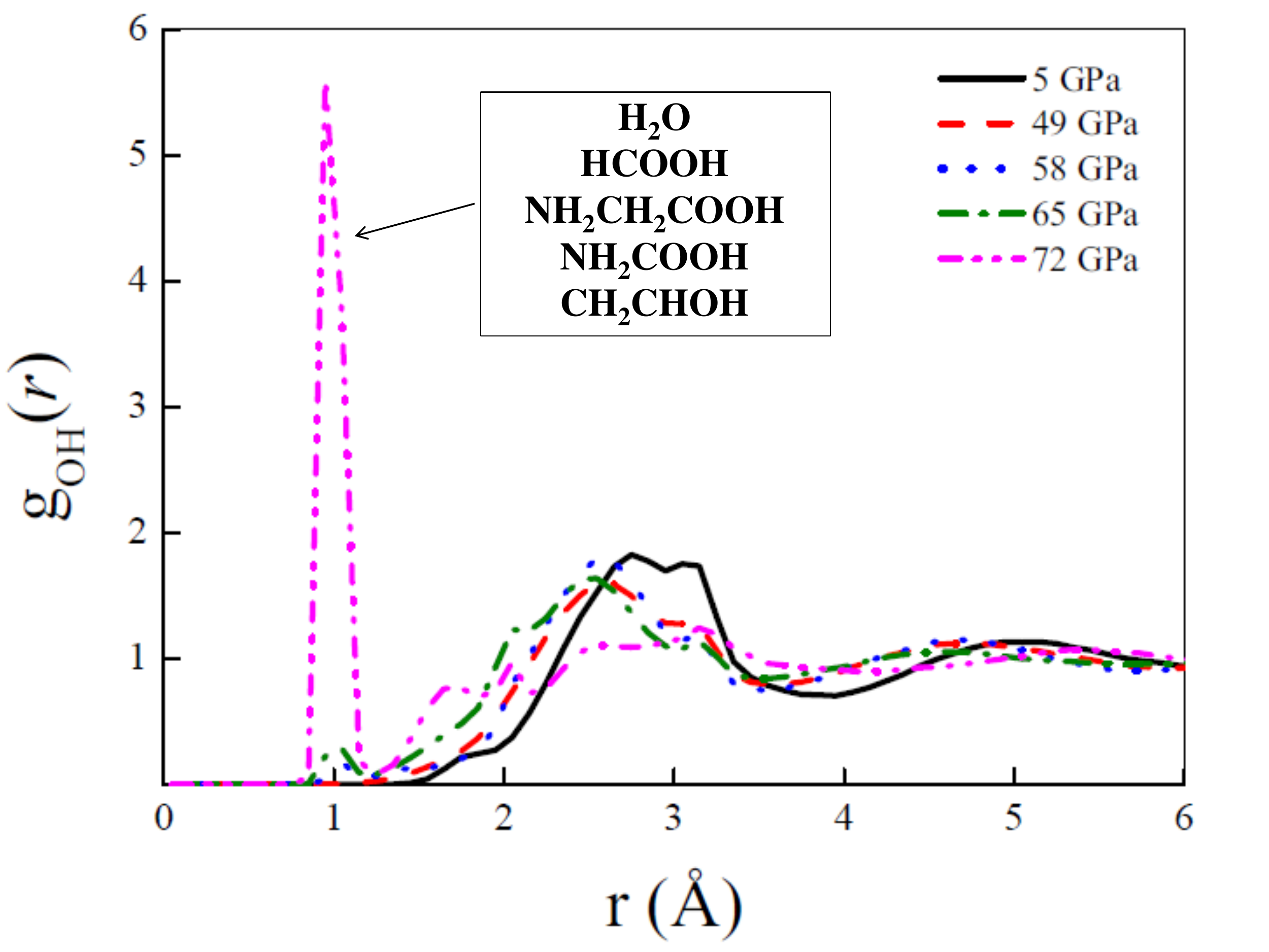} \\
\includegraphics[width=0.4\textwidth]{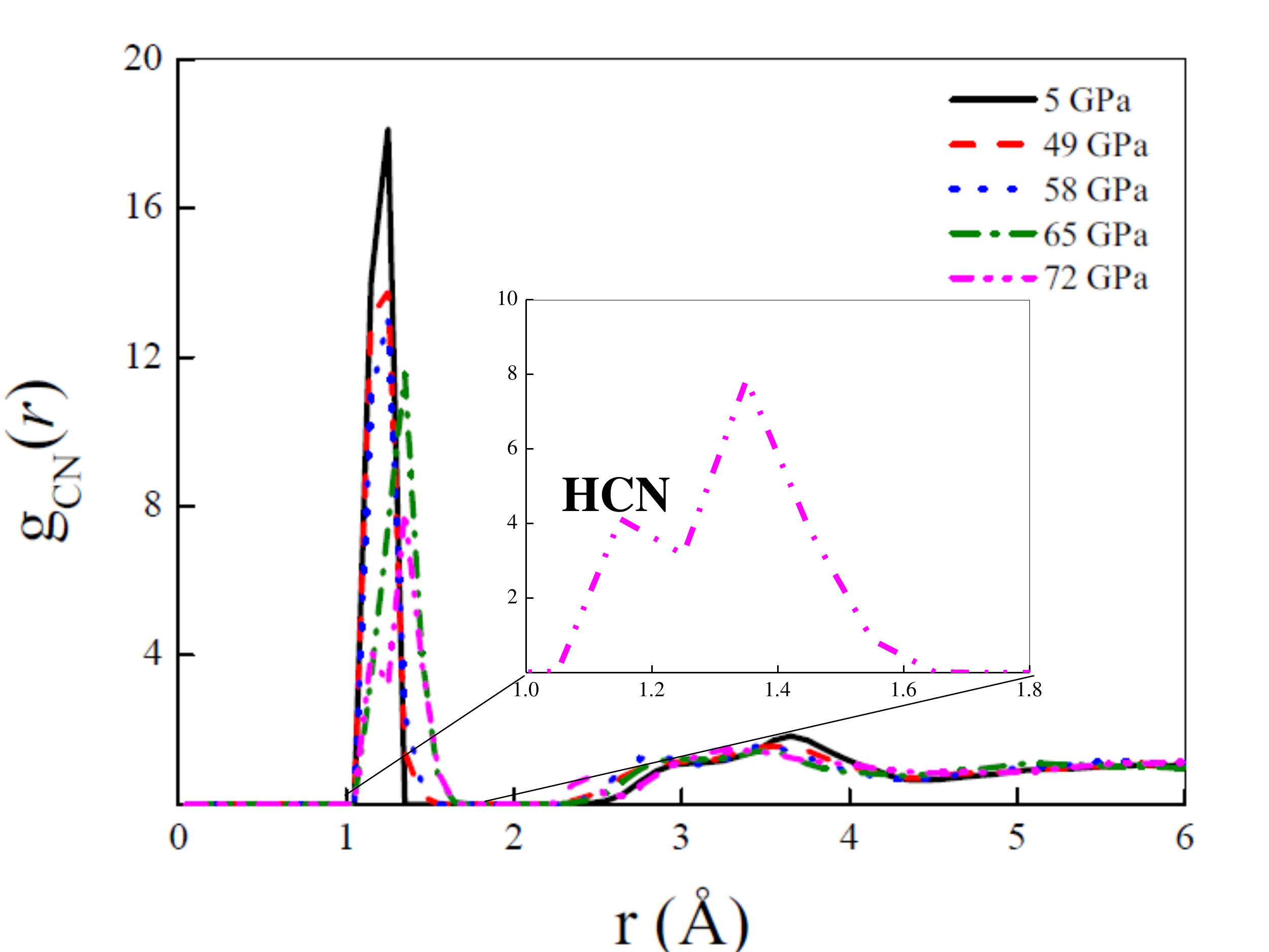} \\
\includegraphics[width=0.4\textwidth]{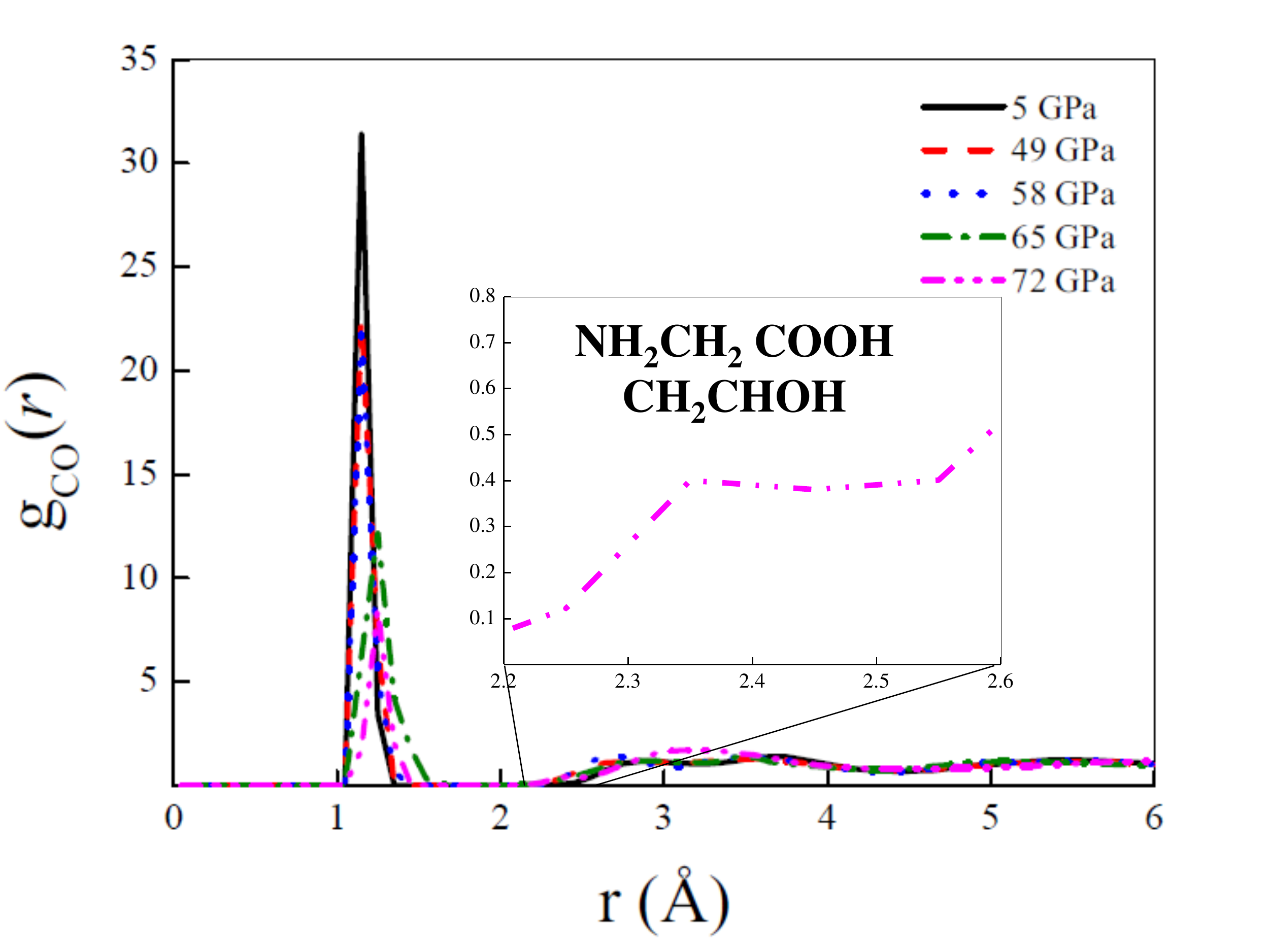} 
\caption{RDFs for oxygen-hydrogen (top panel), carbon-nitrogen (middle panel), and carbon-oxygen (bottom panel) atom pairs. Labels indicate the peak pressure reached at the impact. Whereas the inset of middle panel magnifies the peaks associated with the formation of HCN and CN$^{-}$, the inset of the bottom panel highlights a shoulder in the RDF indicating the onset of glycine and vinyl alcohol at the highest shock velocity.}
\label{ftwo}
\end{figure}

We have not observed any chemical activity when subjecting the mixture of HNCO and H$_2$ to a uni-axial shock wave propagating at $6$~km~s$^{-1}$. Notwithstanding in the highly compressed state ~\textendash~ when pressures reach $37$~GPa (Hugoniot temperature $T_{\rm H} = 857$~K) ~\textendash~ some strong inter-molecular interactions between HNCO molecules are evidenced, once the system is left to relax no reaction products are detected.

The situation changes significantly when the impact velocity exceeds $7$~km~s$^{-1}$, generating peak pressures equal to or higher than $49$~GPa ($T_{\rm H} \geq1213$~K). After decompression we observed formation of formamide and carbamoyl isocyanate (H$_{2}$NCONCO). Although the amount of the newly formed products is relatively low, about $3$\% of the initial HNCO content, it is suggestive that formamide is the first compound synthesized from a sample of HNCO and H$_{2}$. As shown in Figure~\ref{fone}, the first peak of the hydrogen-hydrogen RDF clearly decreases with increasing pressure.  This indicates that the amount of H$_2$ converted into more complex species increases when the system is subjected to progressively more intense shock impacts. It is interesting to note that in shocks impacting on interstellar and circumstellar regions, the formation of species relying on endothermic reactions is enhanced by several orders of magnitude as the shock velocities increases to over 7~km~s$^{-1}$ \citep{LS13}. Furthermore, similar impact velocities are required to explain the observed excitation of the pure rotational lines of H$_2$ (e.g., \citealt{G14}). At the maximum considered pressure ($P_S = 72$~GPa, $v_{\rm d} = 10$~km~s$^{-1}$), after decompression, only $4$\% of the initial amount of H$_2$ remains in unreacted form, which is reflected in the RDFs (see Figure ~\ref{fone}) by a reduction of the intensity of the typical sharp first peak located at $0.76$~{\AA}.

In Table~\ref{tone} we provide with an inventory of the species formed in our simulations performed assuming various uni-axial shock wave velocities. A shock pressure of $58$~GPa ($T_{\rm H} = 1550$~K) induces the formation of formamide and carbamoyl isocyanate, this time each product representing about the $6$\% of the original amount of HNCO (twice the case of $P_S = 49$~GPa). At $P_S = 65$~GPa (or $v_{\rm d} = 9$~km~s$^{-1}$), the amount of synthesized formamide jumps to $16$\% of the original HNCO content.  At this pressure other products, such as formylurea (H$_{2}$NCONHCHO, 3\%) and allophanate (H$_{2}$NCONHCOO$^{-}$, 3\%), start to form in detectable concentration, together with C-N containing aliphatic chains which exhibit carbamate (H$_{2}$NCOO$^{-}$) and formylazanium (NH$_{3}$CHO$^{+}$) groups.
\renewcommand{\tabcolsep}{0.45cm}
\begin{table*}
\caption{Inventory of species formed in the impact simulations under various conditions. $P_S$: maximum pressure reached in the simulation; $T_{\rm H}$: maximum temperature reached; $v_{\rm d}$: velocity of the uni-axial shock wave.}  
\center
\begin{tabular}{cccccc}
\hline \hline
\multicolumn{6}{c}{$P_S$ (GPa) $\vert$ $T_{\rm H}$ (K) $\vert$ $timescale$ (km~s$^{-1}$)} \\
\hline
\renewcommand{\arraystretch}{1}
5 $\vert$ - $\vert$ 0 & 37 $\vert$ 857 $\vert$ 6 & 49 $\vert$ 1213 $\vert$ 7 & 58 $\vert$ 1550 $\vert$ 8 & 65 $\vert$ 1837 $\vert$ 9 & 72 $\vert$ 2122 $\vert$ 10 \\
\hline
H$_{2}$ & H$_{2}$ & H$_{2}$       & H$_{2}$         & H$_{2}$               & H$_{2}$  \\
HNCO    & HNCO    & HNCO          & HNCO            & HNCO                  & CO$_{2}$           \\
        &         & H$_{2}$NCHO   & H$_{2}$NCHO     & H$_{2}$NCHO           & H$_{2}$O \\
        &         & H$_{2}$NCONCO & H$_{2}$NCONCO   & H$_{2}$NCONHCHO       & NH$_{3}$, NH$_{4}^{+}$      \\
        &         &               &                 & H$_{2}$NCONHCOO$^{-}$ & HCN, CN$^{-}$                 \\
        &         &               &                 & H$_{2}$NCOO$^{-}$, H$_{3}$NCHO$^{+}$ complex & HCOOH \\
        &         &               &                 &                                        & H$_{2}$NCHO  \\
        &         &               &                 &                                        & CH$_{2}$CHOH \\
        &         &               &                 &                                        & H$_{2}$NCOOH \\
        &         &               &                 &                                        & CH$_{2}$NH   \\
        &         &               &                 &                                        & CH$_{3}$NH$_{2}$\\
        &         &               &                 &                                        & NH$_{2}$CH$_{2}$COOH\\
        &         &               &                 &                                        & CH$_{3}$CHNH \\
        &         &               &                 &                                        & CH$_{2}$CHNH$_{2}$ \\
        &         &               &                 &                                 & HNC(NH$_{2}$)$_{2}$ complex\\
        &         &               &                 &                                 & C-N aliphatic chains  \\
\hline \hline
\end{tabular}
\label{tone}
\end{table*}

As shown in Figure~\ref{ftwo} (top panel), upon reaching the shock pressure of $65$~GPa an intra-molecular peak rises in the oxygen-hydrogen RDF at ca.~1~{\AA}, which becomes absolutely dominant at $72$~GPa. The peak can be assigned to the formation of  water (38\%), formic acid (HCOOH, 3\%), glycine (NH$_{2}$CH$_{2}$COOH, 3\%), carbamic acid (NH$_{2}$COOH, 3\%), and vinyl alcohol (CH$_{2}$CHOH, 3\%). Remarkably, this is accompanied with the onset of a new peak at $1.16$~{\AA} on the carbon-nitrogen RDF above $65$~GPa (Figure \ref{ftwo}, middle panel) which is associated with the formation of HCN and CN$^{-}$ species. At the highest shock pressures investigated, a shoulder in the carbon-oxygen RDF at about $2.35$~{\AA} suggests formation of glycine and vinyl alcohol (Figure \ref{ftwo}, bottom panel). 
\begin{figure}
\includegraphics[width=0.4\textwidth]{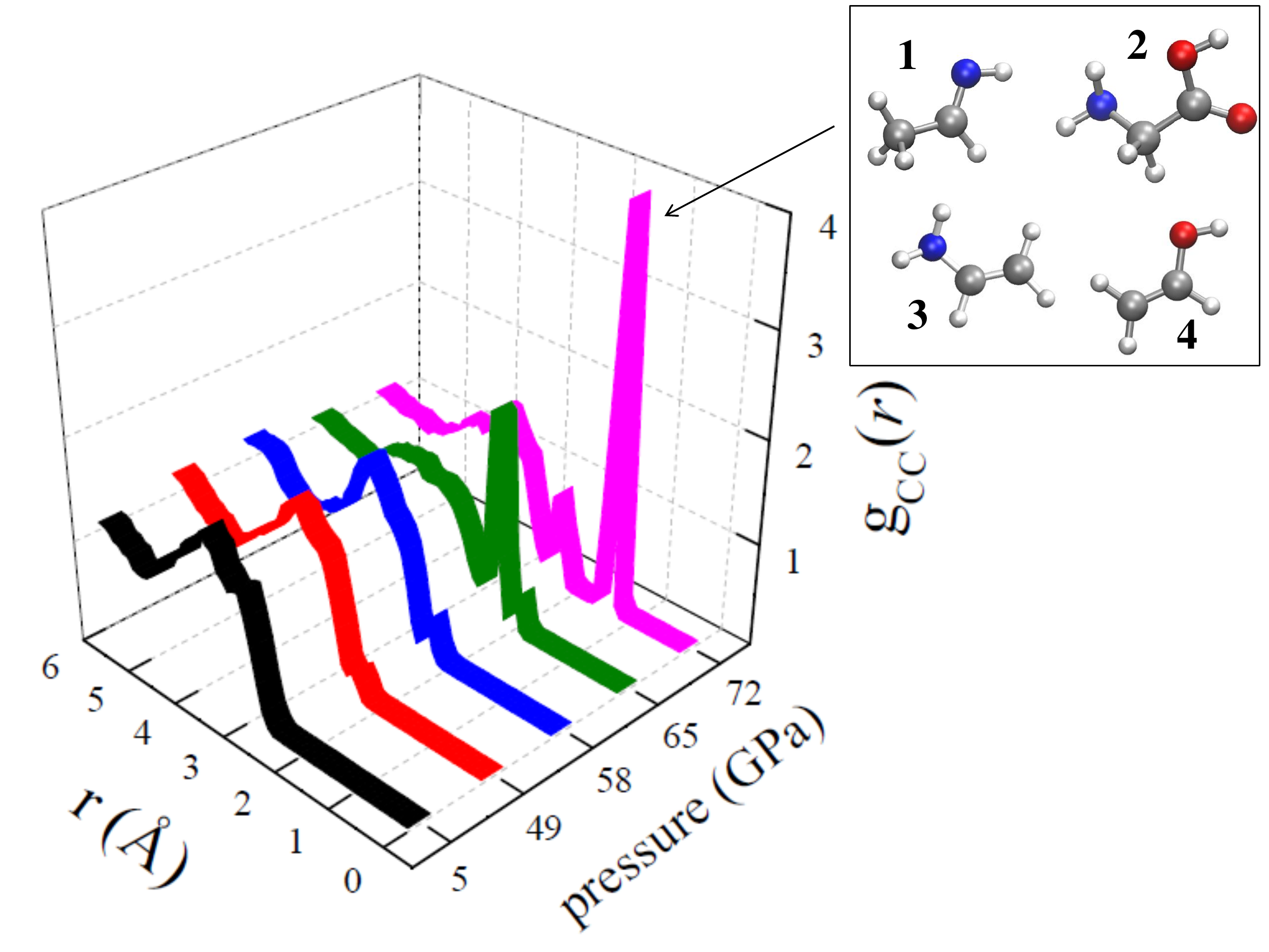}
\caption{Carbon-carbon RDFs for different shock pressures and determined after decompression of the simulation boxes. The onset of the peak at $1.35$~{\AA} is due to the presence of ethanimine (1), glycine (2), vinylamine (3), and vinyl alcohol (4) (see inset).}
\label{fthree}
\end{figure}
The fact that many species exhibit similar yields is due to the statistically exiguous
number of molecules of the starting numerical sample which is, however, typical of high-demanding \emph{ab initio} molecular dynamics simulations. 

The presence of some C-N containing aliphatic chains is underlined by the onset of novel first peaks in the carbon-carbon RDF characterizing the sample subjected to a shock wave producing a maximum pressure of $65$~GPa. In particular, as shown in Figure~\ref{fthree}, two first peaks rise up in the relative C-C RDF, whereas just a small first peak can be recognized in the C-C RDF for shock pressures equal to $49$~GPa and $58$~GPa (which, in such cases, is ascribable to the presence of carbamoyl isocyanate). In addition, as listed in Table~1, also an interesting guanidine-containing complex has been synthesized at $72$~GPa.

Finally, a striking result is the onset of many C-C covalent bonds once the system is relaxed to standard pressures after being shock-compressed up to $72$~GPa by a shock wave propagating at $10$~km~s$^{-1}$, as testified by the impressive first peak located at $1.35$~{\AA} in Figure~\ref{fthree}, which represents the signature of the birth of new C-C bonded species in the sample. In addition to the already mentioned glycine and vinyl alcohol, also syntheses of ethanimine (CH$_{3}$CHNH, 3\%) and its tautomeric form, vinylamine (CH$_{2}$CHNH$_{2}$, 3\%) have been observed. Whereas the former is the building block of alanine, the latter is one of the major components of a vast series of amino-acids: asparagine, aspartic acid, cysteine, leucine, phenylalanine, serine, and tyrosine. 

\begin{figure*}
\centering
\includegraphics[width=0.7\textwidth]{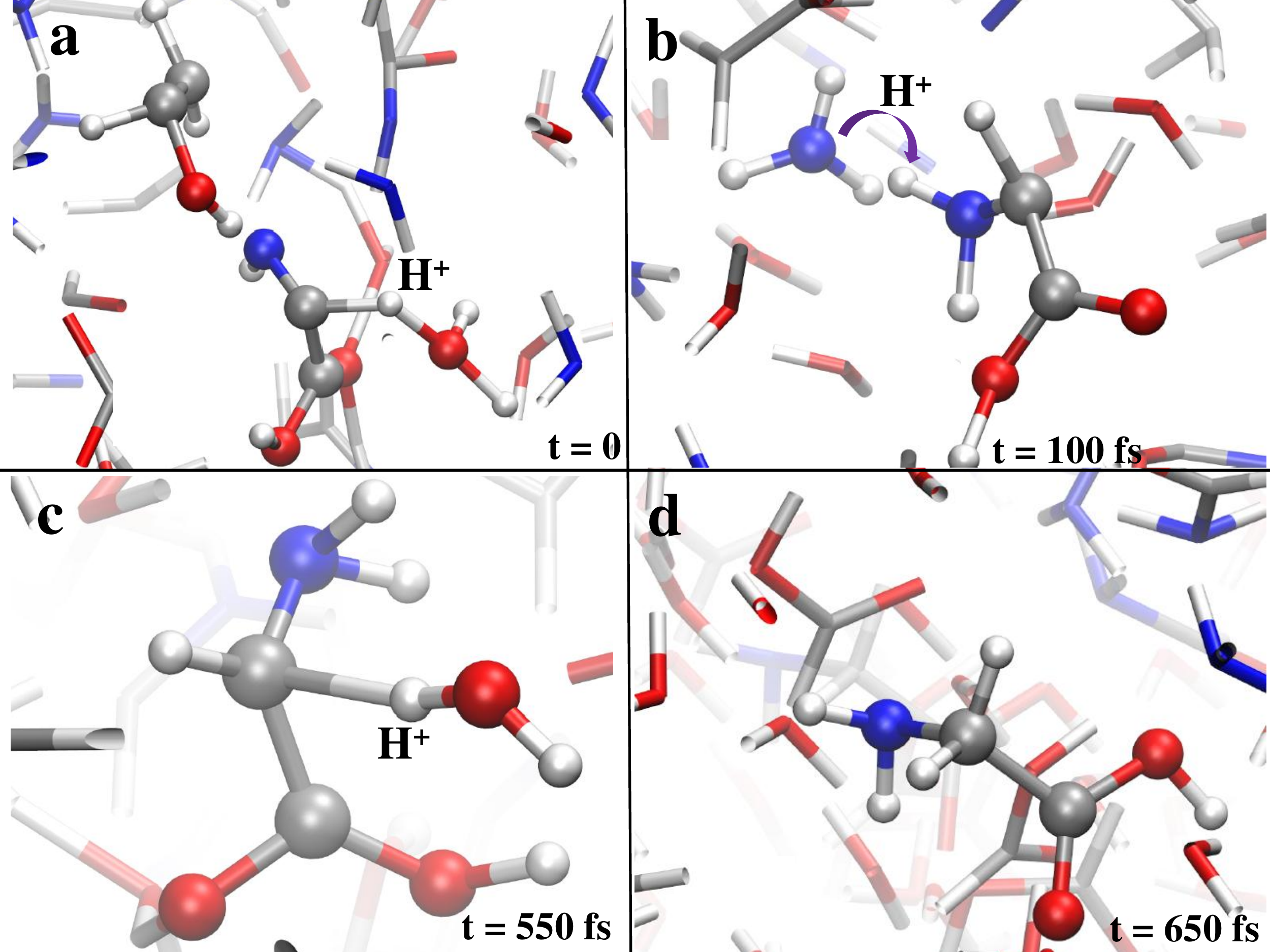}
\caption{Glycine formation mechanism during the decompression of an HNCO and H$_2$ mixture where a shock wave propagating at $10$~km s$^{-1}$ impacted. Red, silver, blue, and white coloring refers to oxygen, carbon, nitrogen, and hydrogen atoms, respectively. After the compression of the simulation cell, where a pseudo-glycine-containing complex has been observed, an highly ionized glycine anion is transiently detected. After successive proton transfer events due to an hydronium cation (a), an ammonium cation (b), and a water molecule (c), glycine is stably synthesized (d).}
\label{ffive}
\end{figure*}

Upon intense compression, a molecular system typically achieves transient inter-molecular distances that instantaneously generates the formation of exotic, short-lived, polymers and complexes. At $P_S = 72$~GPa a pseudo-glycine-containing complex has been observed, partially resembling those detected by \citet{G10}. However, a stable glycine is synthesized in our sample by following the reaction pathway depicted in Figure~\ref{ffive}. In particular, soon after the beginning of the decompression process a short-lived ($100$~fs) anion receives a proton, initially from H$_3$O$^+$, and then from ammonium ion NH$_4$$^+$. In a few hundreds of fs the glycine anion is neutralized by a further proton transfer, occurring through a nearby water molecule (which, in turn, will be rapidly neutralized), eventually leading to a stable glycine molecule.     

One potential problem in our description might be the destruction of the ice layer upon impact. Shattering occurs on mechanical timescales, the latter of the order of the collision time. To derive such a time, we use the \citet{R06} estimate for collisions without adhesion $t_{\rm c} \approx 5 \times (c_s/v_{\rm d})^{0.2} (A/c_s)$ \citep{C93}, with $c_s$ the sound speed, and $A = a_1a_2/(a_1+a_2)$ the reduced radius in the impact. Since the speed of sound in ice is nearly 3 times faster than in water, i.e. $c_s \approx 4$~km~s$^{-1}$, considering the impact between particles of the same size $a$, we obtain $t_{\rm c} \approx 825 \times (a/1~\mu{\rm m}) (v_{\rm d}/1~{\rm km~s^{-1}})^{-0.2}$~ps. For an impact velocity $v_{\rm d} = 10$~km~s$^{-1}$, 1~$\mu$m-sized grains experience a collision time $t_{\rm c} \approx 550$~ps, about three orders of magnitude longer than e.g., the formation time of glycine. Since chemical reaction times are of the order of fractions of ps, mechanical destruction of the ice occurs generally when chemical products are stabilized for a very long time.

H$_2$ is trapped in micropores of CO \citep{MC10}, or water \citep{R91}. \citet{MC10} found that the early desorption of CO (from 15 to 23~K) is caused by the release of H$_2$ molecules from the CO ice. Thus, at least locally, the abundance of H$_2$ molecules relative to CO is substantial. As stated in the previous Section, in a future work we shall investigate the role of CO in the ices in the chemistry of grain-grain collisions. 

\section{Conclusions}
We present a new scheme for the synthesis of formamide, glycine, and amino-acid precursors from shock-waves-induced chemistry of systems composed by isocyanic acid and molecular hydrogen. Our quantum-based investigations indicate that mutually colliding dust grains covered by icy layers are able to produce, at the shock wave velocity threshold triggering the chemical evolution of the sample, formamide which, in turn, will be progressively synthesized under the effect of more intense shock compressions. In addition, when shock waves propagating at $10$~km~s$^{-1}$ impact the sample, the simplest amino-acid glycine is spontaneously formed along with ethanimine and vinylamine. The former is the main constituent of alanine whereas the latter is one of the building blocks of seven distinct amino-acids.
The mechanism we propose may be at work in regions in which the level of turbulence is relatively high, such as turbulent diffuse molecular clouds (see e.g., the detection of formamide by \citealt{T17}), and protoplanetary regions. In pre-stellar cores, where the level of turbulence is small ($\la 1$~km~s$^{-1}$), collisions are not energetic enough to induce pressure-driven formation of complex organics. However, impacts are still able to provide impulsive heating of the colliding particles, enough to fuel rapid exothermic chemical reactions, leading to a thermal runaway. Such reactions may involve free radicals, whose mobility driven by the warm-up may liberate sufficient energy to explode \citep{G76}.

In conclusion, in this work we show that conversion from chemical simplicity to chemical complexity can occur very rapidly within the transient events following catastrophic impacts. The proposed mechanism is general and not specific to any single source. Although limited to a particular reactive system, the ubiquity of collisions driven by turbulent motions does indicate that there are astronomical consequences from this idea. One important point is that systems as the one we consider in this work may be the subject of laboratory validation (see e.g., \citealt{F18}). 

\section*{Acknowledgements}
We would like to thank the anonymous referee for comments and suggestions that helped  clarify and improve the manuscript.

G.~C., J.~E.~S., and M.~F. acknowledge the GACR (Czech Science Foundations) under the grant number 17-05076S. 

This work has been supported by the project PRIN-INAF 2016 The Cradle of Life - GENESIS-SKA (General Conditions in Early Planetary Systems for the rise of life with SKA).

We gratefully acknowledge Dr. Mu\~{no}z Caro and Prof. Chen for the enlightening discussion on CO ices.

{}
\end{document}